\documentclass[11pt]{article}
\usepackage{amssymb,amsmath,amsfonts}
\usepackage{graphicx}
\usepackage{graphics}
\usepackage{eepic,epsfig}

\begin{document}

\title{Smith-Purcell radiation on a surface wave}
\author{A. A. Saharian\thanks{%
E-mail: saharian@ysu.am } \\
\textit{Institute of Applied Problems of Physics} \\
\textit{25 Nersessian Street, 0014 Yerevan, Armenia}}
\maketitle

\begin{abstract}
We consider the radiation from an electron in flight over a surface wave of
an arbitrary profile excited in a plane interface. For an electron bunch the
conditions are specified under which the overall radiation essentially
exceeds the incoherent part. It is shown that the radiation from the bunch
with asymmetric density distribution of electrons in the longitudinal
direction is partially coherent for waves with wavelengths much shorter than
the characteristic longitudinal size of the bunch.
\end{abstract}

\bigskip

\begin{center} {\small Talk presented at the International Conference \\
"Electrons, Positrons, Neutrons and X-rays Scattering Under
External Influences"\\ Yerevan-Meghri, Armenia, October 26-30,
2009 }
\end{center}

\bigskip

\section{Introduction}

Surface waves have wide applications in various fields of science and
technology. In the present talk, based on \cite{Mkrt89}-\cite{Mkrt01}, we
discuss the radiation from a charged particle flying over surface acoustic
wave. The physics of this phenomenon is similar to that for the radiation of
particle flying over a diffraction grating (Smith-Purcell radiation). The
latter is used for the generation of the radiation in the range of
millimeter and submillimeter waves.

\section{Radiation from a single electron}

Let a charge $q$ move with constant velocity $\mathbf{v}$ parallel to a
surface wave excited in the plane interface between homogeneous media with
permittivities $\varepsilon _{0}$ and $\varepsilon _{1}$ (see figure \ref%
{fig1}). If the axis $z$ is aligned with the particle trajectory, then the
equation of the interface has the form%
\begin{equation}
x=x_{0}(z,t)=-d+f(k_{0}z\mp \omega _{0}t),  \label{Profile}
\end{equation}%
where $k_{0}$ and $\omega _{0}$ are the wave number and cyclic frequency, $d$
is the distance from the non-excited interface, and $f(u)$ is the function
describing the wave profile, $f(u+2\pi )=f(u)$. We assume that the particle
moves in the medium with permittivity $\varepsilon _{0}$ and will consider
the radiation in this medium.

\begin{figure}[tbph]
\begin{center}
\epsfig{figure=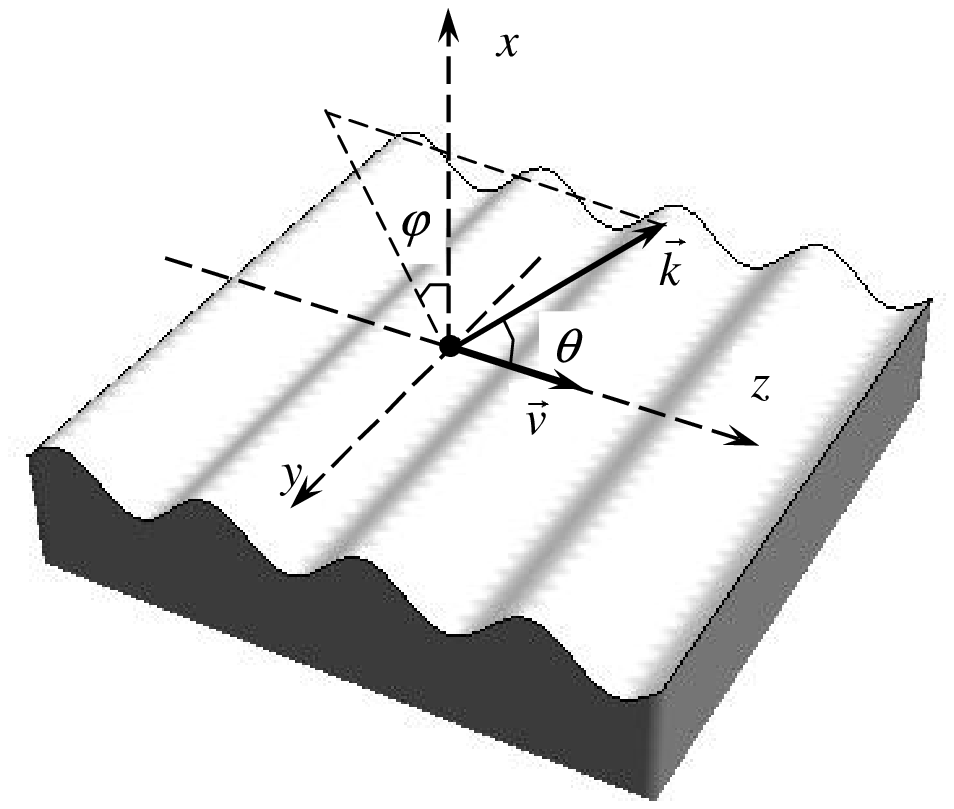,width=8.cm,height=6.cm}
\end{center}
\caption{Electron flying over surface wave excited on the plane interface
between homogeneous media.}
\label{fig1}
\end{figure}

From the symmetry properties of the problem it follows that the emission
angle $\theta $ (with respect to the particle velocity $\mathbf{v}$) and the
frequency $\omega $ of the emitted photon are related by the formula%
\begin{equation}
\omega =\frac{m(k_{0}v\mp \omega _{0})}{1-\beta \sqrt{\varepsilon _{0}}\cos
\theta },\;\beta =\frac{v}{c},  \label{Freq}
\end{equation}%
where $m$ is an integer. The dependence of the radiation intensity on the
distance of the electron trajectory from the non-excited interface, $d$, is
determined by the factor $\exp (-2\omega d{\mathrm{Re\,}}\sigma /v)$, where%
\begin{equation}
\sigma =\left[ (1-\beta ^{2}\varepsilon _{0})\omega _{1}^{2}/\omega
^{2}+\beta ^{2}\varepsilon _{0}\sin ^{2}\theta \sin ^{2}\varphi \right]
^{1/2},\;\omega _{1}=\omega \pm m\omega _{0},  \label{sigma0}
\end{equation}%
with $\varphi $ being the polar angle with respect to the $x$ axis in the
plane perpendicular to the particle trajectory. This factor does not depend
on the specific form of the wave profile in (\ref{Profile}) and is
determined by the dependence of the field spectral components for a
uniformly moving particle on distance $d$. For relativistic particles with $%
\beta ^{2}\varepsilon _{0}\lesssim 1$, one has $\omega _{1}\approx \omega $
and the radiation at large azimuthal angles is exponentially suppressed and
the radiation distribution is strongly anisotropic. For the radiation in the
vacuum at $\varphi \ll 1$ the factor is equal to $\exp (-2\omega
_{1}dv/\gamma )$, with $\gamma =1/\sqrt{1-\beta ^{2}}$ been the Lorentz
factor. In the case $\beta \sqrt{\varepsilon _{0}}<1$, the quantity $\sigma $
is real and the intensity exponentially decreases with increasing distance.
The same is the case for the directions of radiation satisfying the
condition $\sin ^{2}\theta \sin ^{2}\varphi >1-1/(\beta ^{2}\varepsilon _{0})
$, when $\beta \sqrt{\varepsilon _{0}}>1$. For $\beta \sqrt{\varepsilon _{0}}%
>1$ and $\sin ^{2}\theta \sin ^{2}\varphi <1-1/(\beta ^{2}\varepsilon _{0})$%
, the radiation intensity does not depend on particle distance from a
surface of the periodic structure in the absence of absorption. This
corresponds to the reflection of Cherenkov radiation emitted in the first
medium.

In order to determine the radiation intensity we have used two independent
approximate methods. In the first one it is assumed that $|\varepsilon
_{1}-\varepsilon _{0}|\ll \varepsilon _{0}$. The second method, which is
more appropriate for the problem under consideration, assumes that the
amplitude of the surface wave is small. Assuming that the charge moves in
the vacuum $\varepsilon _{0}=1$, and under the condition $\beta \sqrt{%
\varepsilon _{1}-1}>1$, the spectral-angular distribution of the radiation
intensity (per unit path length) in the region $x>0$ is given by the
expression
\begin{eqnarray}
&& \frac{dW}{d\omega d\Omega } =\frac{2q^{2}(\varepsilon
_{1}-1)}{\pi c^{2}v^{2}}\sum_{m\neq 0}\omega ^{3}\sin ^{2}\theta
\cos ^{2}\varphi
|f_{m}|^{2}e^{-2\omega \sigma d/v}  \notag \\
&& \qquad \times \frac{A_{1}\sigma ^{2}+A_{2}\sin ^{2}\theta \sin
^{2}\varphi +A_{3}\beta ^{2}\sin ^{4}\theta \sin ^{4}\varphi
}{\left( 1+\beta ^{2}\sin ^{2}\theta \sin ^{2}\varphi \right)
(\sigma _{2}+\varepsilon _{1}\sin \theta
\cos \varphi )^{2}}\delta \left( \cos \theta -\frac{1}{\beta }+\frac{mk_{0}c%
}{\omega }\right) ,  \label{dW}
\end{eqnarray}%
where $d\Omega =\sin \theta d\theta d\varphi $, and we have assumed that $%
v\gg \omega _{0}/k_{0}$. In (\ref{dW}),  $f_{m}=\frac{1}{2\pi }\int_{-\pi
}^{+\pi }du\,f(u)e^{-imu}$ is the Fourier transform of the profile function,%
\begin{equation}
\sigma _{1}=\sqrt{\beta ^{2}(\varepsilon _{1}-\sin ^{2}\theta \sin
^{2}\varphi )-1},  \qquad  \sigma _{2}=\sqrt{\varepsilon
_{1}-1+\sin ^{2}\theta \sin ^{2}\varphi }, \label{sigma}
\end{equation}%
and
\begin{eqnarray}
A_{1} &=&[\sigma _{1}\sigma _{2}\cos \theta -\varepsilon _{1}(1-\sin
^{2}\theta \cos ^{2}\varphi )]^{2}  \notag \\
&&+\sin ^{2}\theta \lbrack \sigma _{1}(\sigma _{2}\cos \varphi +\sin \theta
\sin ^{2}\varphi )-\varepsilon _{1}\cos \theta \cos \varphi ]^{2},  \notag \\
A_{2} &=&\sigma ^{2}(\sigma _{1}\cos \theta +\varepsilon _{1}\sin \theta
\cos \varphi )^{2}  \notag \\
&&+\beta ^{2}(\sigma _{2}\sin \theta \cos \varphi +\beta \cos \theta \sin
^{2}\theta \sin ^{2}\varphi +\cos ^{2}\theta )^{2},  \notag \\
A_{3} &=&\sigma _{2}^{2}(1-\beta \cos \theta )^{2}+[\cos \theta +\beta \sin
\theta (\sigma _{2}\cos \varphi +\sin \theta \sin ^{2}\varphi )]^{2}.
\label{A123}
\end{eqnarray}%
For a sinusoidal surface wave, $f(u)=a\sin u$, one has $f_{m}=\pm
(a/2i)\delta _{m,\pm 1}$.

\begin{figure}[tbph]
\begin{center}
\epsfig{figure=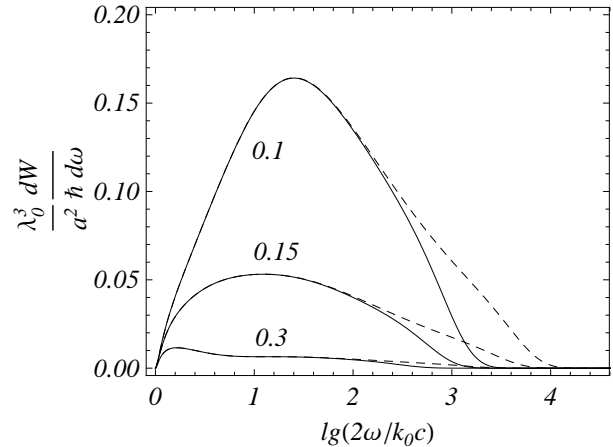,width=8.cm,height=6.cm}
\end{center}
\caption{Spectral density of radiation intensity as a function of the
frequency of the radiated photon. The numbers near the curves correspond to
the values of the ratio $d/\protect\lambda _{0}$.}
\label{fig2}
\end{figure}

\begin{figure}[tbph]
\begin{center}
\epsfig{figure=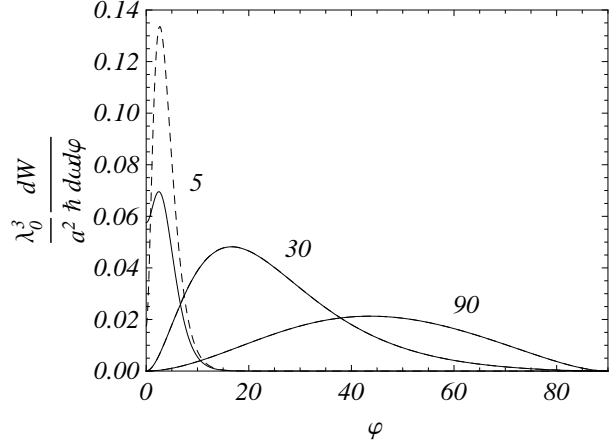,width=8.cm,height=6.cm}
\end{center}
\caption{Spectral-angular density of radiation intensity as a function of
the azimuthal angle of the radiated photon for various values of $\protect%
\theta $.}
\label{fig3}
\end{figure}

We have numerically evaluated the spectral-angular distribution of the
radiation intensity for various values of the parameters in the case of
sinusoidal profile. In this case the only contribution to the radiation
intensity comes from the harmonic $m=1$. The results of these calculations
show that the parameters of the radiation may be effectively controlled by
tuning the characteristics of the surface wave. For an illustration we plot
in figure \ref{fig2} the spectral distribution of the radiation intensity
\begin{equation}
\frac{dW}{d\omega }=\int_{0}^{\pi }d\theta \,\sin \theta \int_{-\pi /2}^{\pi
/2}d\varphi \frac{dW}{d\omega d\Omega },  \label{IntInteg}
\end{equation}%
in the case of a sinusoidal surface wave as a function of the frequency of
the radiated photon for various values of the ratio $d/\lambda _{0}$
(numbers near the curves), where $\lambda _{0}=2\pi /k_{0}$ is the
wavelength of the surface wave. The full (dashed) curves correspond to the
electron energy $E_{e}=100$ MeV ($E_{e}=500$ MeV). Note that $\omega
/k_{0}c=\lambda _{0}/\lambda $ with $\lambda $ being the wavelength of the
radiated photon. The corresponding radiation angle is related to the
frequency by the relation $\cos \theta =1/\beta -k_{0}c/\omega $. In
accordance with this relation, for relativistic electrons, large values of $%
\omega /(k_{0}c)$ correspond to small angles $\theta $. For $\theta \gg
\gamma ^{-1}$ the radiation intensity is relatively insensitive to the
particle energy, whereas for $\theta \lesssim \gamma ^{-1}$ the intensity
strongly increases with increasing energy. From figure \ref{fig2} the
suppression of the radiation intensity with increasing $d$ is well seen. In
figure \ref{fig3} we present the radiation intensity
\begin{equation}
\frac{dW}{d\omega d\varphi }=\int_{0}^{\pi }d\theta \,\sin \theta \frac{dW}{%
d\omega d\Omega },  \label{Intens}
\end{equation}%
as a function of the azimuthal angle $\varphi $ for various values of $%
\theta $ (numbers near the curves) and for $d/\lambda _{0}=0.15$.
As in the case of figure \ref{fig1}, the full (dashed) curves
correspond to the electron energy $E_{e}=100$ MeV ($E_{e}=500$
MeV). As we have mentioned, for $\theta \gg \gamma ^{-1}$ the
dependence of the radiation intensity on the particle energy is
weak and for $\theta =30^{\circ },90^{\circ }$ the full and dashed
curves coincide. For small angles $\theta $ the radiation is
emitted mainly along $\varphi \lesssim \gamma ^{-1}$.

\section{Radiation from an electron bunch}

In this section, as a source of the radiation we shall consider a cold bunch
consisting of $N$ electrons and moving with constant velocity $v$ along the $%
z$ axis (for coherence effects in the Smith-Purcell radiation see also \cite%
{Mkrt98}). The density of the current in the bunch can be written in the form%
\begin{equation}
\mathbf{j}=q\mathbf{v}\sum_{j=1}^{N}\delta (\mathbf{r}-\mathbf{R}_{j}-%
\mathbf{v}t),  \label{jbunch}
\end{equation}%
with $\mathbf{R}_{j}=(X_{j},Y_{j},Z_{j})$ being the position of the $j$-th
particle at the initial moment $t=0$. The spectral density of the radiation
energy flux for a given $m$, defined by the relation
\begin{equation}
\frac{c}{4\pi }\int_{-\infty }^{+\infty }dt\left[ \mathbf{EH}\right]
=\int_{0}^{\infty }d\omega \sum_{m}\mathbf{P}_{m}^{\left( N\right) }\left(
\omega \right) ,  \label{IntensVec}
\end{equation}
with $\mathbf{E}$ and $\mathbf{H}$ being the electric and magnetic fields,
can be written in the form%
\begin{equation}
\mathbf{P}_{m}^{(N)}(\omega )=\mathbf{P}_{m}^{(1)}(\omega )S_{N}\,.
\label{PmN}
\end{equation}%
In (\ref{PmN}), $\mathbf{P}_{m}^{(1)}(\omega )$ is the corresponding
function for the radiation from a single electron with coordinates $x=y=z=0$
at the initial moment $t=0$, and%
\begin{equation}
S_{N}=\left\vert \sum_{j=1}^{N}\exp \left[ -\omega _{1}\sigma
X_{j}/v-ik_{y}Y_{j}-i\omega _{1}Z_{j}/v\right] \right\vert ^{2}.  \label{Sn}
\end{equation}%
Assuming that the coordinates of the $j$-th particle are independent random
variables and averaging the quantity (\ref{Sn}) over the positions of a
particle in the bunch, we obtain%
\begin{equation}
\left\langle \mathbf{P}_{m}^{(N)}(\omega )\right\rangle =\mathbf{P}%
_{m}^{(1)}(\omega )\left\langle S_{N}\right\rangle ,\;\left\langle
S_{N}\right\rangle =Nh+N(N-1)|h_{x}h_{y}h_{z}|^{2},  \label{PNmij}
\end{equation}%
where we have introduced the notations%
\begin{eqnarray}
h &=&\left\langle \exp (2\omega X{\mathrm{Re\,}}\sigma /v)\right\rangle
,\;h_{l}=\left\langle \exp (iK_{l}l)\right\rangle ,\;l=x,y,z,  \notag \\
K_{x} &=&-i\omega \sigma /v,\;K_{y}=k_{y}=(\omega \sqrt{\varepsilon _{0}}%
/c)\sin \theta \sin \varphi ,\;K_{z}=\omega _{1}/v.  \label{Kz}
\end{eqnarray}%
The functions $|h_{l}|^{2}$ determine the bunch form factors in the
corresponding directions. In (\ref{PNmij}), the term proportional to $N^{2}$
determines the contribution of coherent effects. Conventionally it is
assumed that the coherent radiation is produced at wavelengths equal and
longer than the electron bunch length. However, as we shall see below, this
conclusion depends on the distribution of electrons in the bunch.

As it is seen from (\ref{Kz}), the form factors in $y$ and $z$ directions
are determined by the Fourier transforms of the corresponding bunch
distributions. First let us consider a Gaussian distribution:
\begin{equation}
f_{l}=\frac{1}{\sqrt{2\pi }b_{l}}\exp \left( -\frac{l^{2}}{2b_{l}^{2}}%
\right) ,\quad l=x,y,z,  \label{gauss1}
\end{equation}%
with $b_{x}$, $b_{y}$ and $b_{z}$ being the corresponding characteristic
sizes of the bunch. Assuming that all particles of the bunch are in the
medium with permittivity $\varepsilon _{0}$ and, therefore, $b_{x}<d-a$,
with $a$ being the surface wave amplitude, one finds
\begin{equation}
\langle S_{N}\rangle =N\exp \left[ \frac{2\omega ^{2}}{v^{2}}(\mathrm{Re\,}%
\sigma )^{2}b_{x}^{2}\right] \left[ 1+(N-1)\exp \left( -\frac{\omega ^{2}}{%
v^{2}}\left\vert \sigma \right\vert ^{2}b_{x}^{2}-k_{y}^{2}b_{y}^{2}-\frac{%
\omega _{1}^{2}b_{z}^{2}}{v^{2}}\right) \right] ,  \label{factgauss}
\end{equation}%
where the second term in the square brackets determines the relative
contribution of the coherent effects. For a non-relativistic bunch with $%
b_{y}\gtrsim b_{x,z}$, the corresponding exponent is equal to
\begin{equation*}
\exp [-(2\pi m/\lambda _{0})^{2}(b_{x}^{2}+b_{z}^{2})-(2\pi /\lambda
)^{2}b_{y}^{2}\sin ^{2}\theta \sin ^{2}\varphi ]
\end{equation*}%
(here we consider the case $\varepsilon _{0}=1$). Note that in this case for
the wavelength of the radiation one has $\lambda \sim \lambda _{0}/\beta m$
and the coherent effects are exponentially suppressed when $\lambda \lesssim
b_{l}/\beta $. For a relativistic bunch the relative contribution of
coherent effects is given by
\begin{equation}
N\exp \{-(2\pi /\lambda )^{2}[(b_{x}^{2}+b_{y}^{2})\sin ^{2}\theta \sin
^{2}\varphi +b_{z}^{2}]\},  \label{relativecoh}
\end{equation}%
for $\sin \theta \sin \varphi >\gamma ^{-1}$, and by
\begin{equation}
N\exp [-(2\pi b_{x}/\lambda \gamma )^{2}-(2\pi /\lambda
)^{2}(b_{z}^{2}+b_{y}^{2}\sin ^{2}\theta \sin ^{2}\varphi )],
\label{relativecoh1}
\end{equation}%
for the radiation with $\sin \theta \sin \varphi \lesssim \gamma ^{-1}$. As
we see, in this case the transverse form factor is strongly anisotropic. It
follows from (\ref{factgauss}) that for a real $\sigma $, fixed electron
number and fixed distance of the bunch center from the surface wave the
radiation intensity exponentially increases with increasing $b_{x}$. This is
because the number of electrons passing close to the surface wave increases.
The number of electrons with long distances will also increase. But the
contribution of close electrons surpasses the decrease of the intensity due
to distant ones.

As we have seen, for a Gaussian distribution the relative contribution of
coherent effects is exponentially suppressed in the case $b_{i}>\lambda $.
This result is a consequence of the mathematical fact that for a function $%
f(x)\in C^{\infty }(R)$ one has the estimate
\begin{equation}
F(u)\equiv \int_{-\infty }^{+\infty }f(l)e^{iul}dl=O(u^{-\infty }),\quad
u\rightarrow +\infty ,  \label{factornotat}
\end{equation}%
where $u=2\pi b_{z}/\lambda $ for the longitudinal form factor of the
relativistic bunch.

We have considered the case of Gaussian distribution in the bunch. However,
it should be noted that due to various beam manipulations the bunch shape
can be highly non-Gaussian. In the estimate (\ref{factornotat}) the
continuity condition for the function $f(l)$ and for infinite number of its
derivatives is essential. It can be seen that when $f(l)\in C^{n-1}(R)$, and
the derivative $f^{(n)}(l)$ is discontinuous at point $l_{1}$, we have the
asymptotic estimate
\begin{equation}
F(u)=(-iu)^{-n-1}\left[ f^{(n)}(l_{1}+)-f^{(n)}(l_{1}-)\right] ,\quad
u\rightarrow +\infty .  \label{asympdiscont}
\end{equation}%
Unlike the case of (\ref{factornotat}), now the form factor for the short
wavelengths decreases more slowly, as power-law, $(2\pi b_{l}/\lambda
)^{-n-1}$. In the coherent part of the radiation this form factor is
multiplied by a large number, $N$, of particles per bunch and the coherent
effects dominate under the condition
\begin{equation}
2\pi b_{l}/\lambda <N^{1/2(n+1)}.  \label{cohcondpower}
\end{equation}%
The radiation intensity is enhanced by the factor $N(\lambda /2\pi
b_{l})^{2(n+1)}$. Since conventionally there are $10^{8}-10^{10}$
electrons per bunch, the condition (\ref{cohcondpower}) can be
easily met even in the case $2\pi b_{l}/\lambda >1$. For instance,
in the case of $n=1$, $N\sim 10^{10}$, the coherent radiation
dominates for $100\lambda >b_{l}$.

As an example, let us discuss the case when the electrons are normally
distributed in the $x$ and $y$ directions and the distribution function in
the $z$ direction has an asymmetric Gaussian form
\begin{equation}
f(z)=\frac{2}{\sqrt{2\pi }(1+p)b_{l}}\left[ \exp \left( -\frac{l^{2}}{%
2p^{2}b_{l}^{2}}\right) \theta (-l)+\exp \left( -\frac{l^{2}}{2b_{l}^{2}}%
\right) \theta (l)\right] ,  \label{asgauss}
\end{equation}%
where $\theta \left( l\right) $ is the unit step function, $l_{0}=(1+p)b_{l}$
is the characteristic bunch length, parameter $p$ determines the degree of
bunch asymmetry. Now the expression for $\langle S_{N}\rangle $ can be
written as
\begin{equation}
\langle S_{N}\rangle =N\exp \left( \frac{2\omega ^{2}}{v^{2}}\sigma
^{2}b_{x}^{2}\right) \left[ 1+(N-1)\exp \left( -\frac{\omega ^{2}}{v^{2}}%
\sigma ^{2}b_{x}^{2}-k_{y}^{2}b_{y}^{2}\right) \left\vert F(\omega
_{1}/v)\right\vert ^{2}\right] ,  \label{esen1}
\end{equation}%
where
\begin{equation}
F(u)=\frac{1}{p+1}\left\{ e^{-t^{2}}+pe^{-p^{2}t^{2}}-\frac{2i}{\sqrt{\pi }}%
\left[ W(t)-pW(pt)\right] \right\} ,  \label{factor3}
\end{equation}%
with the notation
\begin{equation}
W(t)=\int_{0}^{t}\exp (l^{2}-t^{2})dl,\quad t=\frac{ub_{l}}{\sqrt{2}}.
\label{notat3}
\end{equation}%
The expression $\left\vert F(u)\right\vert ^{2}$ is invariant with respect
to the replacement $p\rightarrow 1/p$, $b_{l}\rightarrow b_{l}p$ that
corresponds to the mirror reversal of the bunch. When the electron
distribution is symmetric $(p=1)$, the second summand in the square brackets
of (\ref{factor3}) vanishes and, as was mentioned earlier, the form factor
exponentially decreases for short wavelengths $\lambda <2\pi b_{l}/\beta $.
For $pt\gg 1$ the asymptotic behaviour of the function $F(u)$ is found from (%
\ref{factor3}) and has the form
\begin{equation}
F(u)\sim i\sqrt{\frac{2}{\pi }}\frac{1-p}{u^{3}b_{l}^{3}p^{2}}.
\label{asympasymgauss}
\end{equation}

In the case of a relativistic bunch and for $\varphi >\gamma ^{-1}$, the
exponent of the second summand in the square brackets of (\ref{esen1}) is of
the order $(2\pi b_{i}/\lambda )^{2},i=x,y$. When the transverse size of the
bunch is shorter than the radiation wavelength, then the relative
contribution of the coherent effects is $\sim N\left\vert F(\omega
_{1}/v)\right\vert ^{2}$. For an asymmetrical bunch this contribution can be
dominant even in the case when the bunch length is greater than the
radiation wavelength. Indeed, according to (\ref{asympasymgauss}) even for
weakly asymmetrical bunch we have $N\left\vert F\right\vert ^{2}\sim
N(v/\omega _{1}b_{z})^{6}$, and the radiation is coherent for $b_{z}\lesssim
\lambda N^{1/6}/(2\pi )$, where we have taken into account that for an
relativistic bunch $\omega \gg m\omega _{0}$ and therefore $\omega
_{1}\approx \omega $ , as was mentioned above. In the case of the bunch with
$N\sim 10^{10}$ and for $2\pi b_{z}/\lambda \lesssim 10$ the radiation is
coherent. For the radiation $\varphi \lesssim \gamma ^{-1}$, the exponent of
the second summand in square brackets of Eq. (\ref{esen1}) for $\varepsilon
_{0}=1$ is of the order $(2\pi b_{x}/\lambda \gamma )^{2}$. Hence, for a
relativistic bunch the radiation in directions $\varphi \lesssim \gamma ^{-1}
$ can be coherent even in the case when the transverse size of the bunch is
greater than the wavelength. For this it is sufficient to have the
conditions
\begin{equation}
b_{x}\lesssim \frac{\gamma \lambda }{2\pi },\quad b_{z}\lesssim \frac{%
\lambda N^{1/6}}{2\pi }.  \label{cond2}
\end{equation}%
The second of these conditions is written for weakly asymmetrical bunches.
In the case of strong asymmetry the corresponding conditions are less
restrictive: $b_{z}\lesssim \lambda N^{1/2}/2\pi $ for $pb_{z}\ll \lambda $ .

Let $f(l,a)$ be a continuous distribution function depending on the
parameter $a$, and $\lim_{a\rightarrow 0}f(l,a)=f(l)$. The integral $F(u,a)$
for $f(l,a)$ uniformly converges and hence $\lim_{a\rightarrow 0}F(u,a)=F(u)$%
. It follows from here that the estimate presented above is valid for
continuous functions as well if they are sufficiently close to the
corresponding discontinuous function (the corresponding derivative is
sufficiently large, see below). Aiming to illustrate this, we consider the
asymmetric distribution
\begin{equation}
f(z,a_{l},b_{l},l_{0})=\frac{1}{4l_{0}}\left[ th\left( \frac{l+l_{0}}{a_{l}}%
\right) -th\left( \frac{l-l_{0}}{b_{l}}\right) \right] .  \label{asymth}
\end{equation}%
For $l_{0}>a_{l},b_{l}$ this function describes a rectangular bunch having
exponentially decreasing asymmetric tails with characteristic sizes $a_{l}$
and $b_{l}$. In the limit $a_{l},b_{l}\rightarrow 0$ one obtains the
rectangular distribution with the bunch length $2l_{0}$. The explicit
evaluation of the expression (\ref{factornotat}) with the function (\ref%
{asymth}) leads to
\begin{equation}
F(u,a_{l},b_{l},l_{0})=\frac{i}{2ul_{0}}\left( \frac{\overline{a}%
_{l}e^{-iul_{0}}}{\sinh \overline{a}_{l}}-\frac{\overline{b}_{l}e^{iul_{0}}}{%
\sinh \overline{b}_{l}}\right) ,  \label{factorforth}
\end{equation}%
with the notations $\overline{a}_{l}\equiv \pi ua_{l}/2$ and $\overline{b}%
_{l}\equiv \pi ub_{l}/2$. From here it follows that if $\overline{a}_{l}\sim
1$ and $\overline{b}_{l}\sim 1$ then $F\sim (ul_{0})^{-1}$ for $ul_{0}\gg 1$%
. In the limit $a_{l},b_{l}\rightarrow 0$, from (\ref{factorforth}) one
obtains the well known form factor for the rectangular distribution: $F_{{%
\mathrm{rect}}}(u,l_{0})=\sin (ul_{0})/(ul_{0})$. As it is seen, the
rectangular distribution is a good approximation for (\ref{asymth}) if $%
a_{l},b_{l}\ll \lambda $. The main contribution to (\ref{factorforth}) comes
from the bunch tails, i.e. from the parts of bunch with large derivatives of
the distribution function. This is the case for the general case of
distribution function as well: if $ul_{0}\gg 1$ the main contribution comes
from the parts of the bunch where $df/d(l/l_{0})\gtrsim u$ and in this case $%
F(u)\sim 1/u$, $u\rightarrow \infty $. This can be generalized for higher
derivatives as well: if $d^{i}f/d(l/l_{0})^{i}\ll u$, $i=1,...,n-1$, and $%
d^{n}f/d(l/l_{0})^{n}\gtrsim u$ then $F(u)\sim (ul_{0})^{-n}$.

For a relativistic bunch one has $u\sim 2\pi /\lambda $ for the form factors
in $y$ and $z$ directions and the conclusion can be formulated as follows.
If for the distribution function one has
\begin{equation}
\lambda \frac{d^{i}f}{d(l/l_{0})^{i}}\ll 2\pi ,\quad
i=1,...,n-1;\quad \lambda \frac{d^{n}f}{d(l/l_{0})^{n}}\gtrsim
2\pi ,  \label{cohcondn}
\end{equation}%
then the relative contribution of coherent effects into the radiation
intensity is proportional to $N(\lambda /2\pi l_{0})^{2n}$ and the radiation
is partially coherent in the case $\lambda <l_{0}$ but $\lambda >2\pi
l_{0}N^{-1/2n}$, with $l_{0}$ being the characteristic bunch size in the
corresponding direction. In this case the main contribution into the
radiation intensity comes from the parts of the bunch with large derivatives
of the distribution function in the sense of the second condition in (\ref%
{cohcondn}). For example, in the case of asymmetric distribution (\ref%
{asgauss}), when $ub_{l}\gg 1$ and $pub_{l}<1$, the main contribution comes
from the left Gaussian tail with $l<0$. For this tail $df/d(l/b_{l})\sim
u/(pub_{l})\gtrsim u$, at $l\sim pb_{l}$ and therefore $F(u)\sim 1/(ub_{l})$%
. This can be seen directly from the exact relation (\ref{factor3}) as well
by using the asymptotic formula for the function $W(t)$.

\section{Conclusion}

In the present talk we have considered the radiation from a single
electron and from an electron bunch of arbitrary structure flying
over the surface wave excited in a plane interface. For small
amplitudes of the surface wave the spectral-angular distribution
of the radiation intensity from a single electron is given by
expression (\ref{dW}). We have shown that the at large angles with
respect to the electron trajectory the dependence of the radiation
intensity on the electron energy is relatively week, whereas at
small angles the intensity strongly increases with increasing
energy. It is demonstrated that the radiation from a bunch can be
partially coherent in the range of wavelengths much shorter than
the characteristic longitudinal size of the bunch and the main
contribution to the radiation intensity comes from the parts of
the bunch with large derivatives of the distribution function. In
this case for short wavelengths the relative contribution of
coherent effects decreases as a power-law instead of exponentially
decreasing. The corresponding conditions for the distribution
function are specified. The coherent effects lead to an essential
increase in the intensity of the emitted radiation.

The author gratefully acknowledges the organizers of the International
Conference "Electrons, Positrons, Neutrons and X-rays Scattering Under
External Influences", Yerevan-Meghri, October 26-30, 2009, for the financial
support to attend the conference.

\end{document}